\documentclass[a4paper, 12pt,reqno]{article}
\usepackage{amsmath}
\usepackage{amssymb}
\usepackage[latin1]{inputenc}
\usepackage{geometry}
\usepackage{bbm}
\usepackage{fancyhdr}
\usepackage{a4wide}
\usepackage{setspace}

\newcommand*{\cA}{\mathcal{A}}
\newcommand*{\cB}{\mathcal{B}}
\newcommand*{\cC}{\mathcal{C}}
\newcommand*{\C}{\mathbb{C}}
\newcommand*{\N}{\mathbb{N}}
\newcommand*{\R}{\mathbb{R}}

\newcommand{\abs}[1]{\left|#1\right|}
\newcommand{\norm}[1]{\left\|#1\right\|}

\newcommand{\lb}{\left(}

\newcommand{\lab}{\left\langle}

\newcommand{\rb}{\right)}

\newcommand{\rab}{\right\rangle}

\newcommand{\1}{\mathbbm{1}}
\newtheorem{satz}{Satz}[section]
\newtheorem{proposition}[satz]{Proposition}
\newtheorem{theorem}[satz]{Theorem}
\newtheorem{corollary}[satz]{Corollary}
\newtheorem{lemma}[satz]{Lemma}
\newtheorem{remark}[satz]{Remark}
\newtheorem{definition}[satz]{Definition}
\newtheorem{example}[satz]{Example}
\allowdisplaybreaks
\begin{document}
$\,$
\vskip 20mm
\begin{center}
\begin{LARGE}A White Noise Approach to Phase Space Feynman Path Integrals\end{LARGE} \\
\begin{small}
 %\subtitle
\end{small}
\end{center}
\vskip 10mm
\begin{center}
\begin{large}Wolfgang Bock \end{large}   \\
\textit{\begin{small}bock@mathematik.uni-kl.de \\
\end{small} }
\begin{large} Martin Grothaus \end{large} \\   
\textit{\begin{small}grothaus@mathematik.uni-kl.de \\
\end{small} }

\bigskip
\begin{small}
\textit{Functional Analysis and Stochastic Analysis Group, \\
Department of Mathematics, \\
University of Kaiserslautern, 67653 Kaiserslautern, Germany \\}
\end{small}
\end{center}

\begin{abstract}
The concepts of phase space Feynman integrals in White Noise Analysis are established. As an example the harmonic oscillator is treated. The approach perfectly reproduces the right physics. I.e.~, solutions to the Schrödinger equation are obtained and the canonical commutation relations are satisfied. The later can be shown, since we not only construct the integral but rather the Feynman integrand and the corresponding generating functional. 
\end{abstract}

\pagebreak

\section{Introduction}
As an alternative approach to quantum mechanics Feynman introduced the concept of path integrals (\cite{F48,Fe51,FeHi65}), which was developed into an extremely useful tool in many branches of theoretical physics. In this article we develop the concepts for realizing Feynman integrals in phase space in the framework of White Noise Analysis. The phase space Feynman integral for a particle moving from $y_0$ at time $0$ to $y$ at time $t$ under the potential $V$ is given by 
\begin{equation}\label{psfey} 
{\rm N} \int_{x(0)=y_0, x(t)=y} \int \exp\left(\frac{i}{\hbar} \int_0^t p\dot{x}-\frac{p^2}{2} -V(x,p) \, d\tau \right) \prod_{0<\tau<t}  dp(\tau) dx(\tau),\quad \hbar = \frac{h}{2\pi}.
\end{equation}
Here $h$ is Planck's constant, and the integral is thought of being over all position paths with $x(0)=y_0$ and $x(t)=y$ and all momentum paths. The missing restriction on the momentum variable at time $0$ and time $t$ is an immediate consequence of the Heisenberg uncertainty relation, i.e.~ the fact that one can not measure momentum and space variable at the same time. The path integral to the phase space has several advantages. Firstly the semi-classical approximation can be validated easier in a phase space formulation and secondly that quantum mechanics are founded on the phase space, i.e.~ every quantum mechanical observable can be expressed as a function of the space and momentum. A discussion about phase space path integrals can be found in the monograph \cite{AHKM08} and in the references therein.\\
In the last fifty years there have been many approaches for giving a mathematically rigorous meaning to the phase space path integral by using e.g.~ analytic continuation, see \cite{KD82,KD84} or Fresnel integrals \cite{AHKM08,AGM02}.  Here we choose a white noise approach.
White Noise Analysis is a mathematical framework which offers generalizations of concepts from finite-dimensional analysis, like differential operators and Fourier transform to an infinite-dimensional setting. We give a brief introduction to White Noise Analysis in Section 2, for more details see \cite{Hid80,BK95,HKPS93,Ob94,Kuo96}. Of special importance in White Noise Analysis are spaces of generalized functions and their characterizations. In this article we choose the space of Hida distributions, see Section 2.\\
The idea of realizing Feynman integrals within the white noise framework goes back to \cite{HS83}. There the authors used exponentials of quadratic (generalized) functions in order to give meaning to the Feynman integral in configuration space representation
\begin{equation*}
{\rm N}\int_{x(0) =y_0, x(t)=y} \exp\left(\frac{i}{\hbar} S(x) \right) \, \prod_{0<\tau<t} \, dx(\tau) ,\quad \hbar = \frac{h}{2\pi},
\end{equation*}
with the classical action $S(x)= \int_0^t \frac{1}{2} m \dot{x}^2 -V(x)\, d\tau$.
We use these concepts of quadratic actions in White Noise Analysis, which were further developed in \cite{GS98a} to give a rigorous meaning to the Feynman integrand 
\begin{align}\label{integrandpot}
I_V = {\rm Nexp}\left( \frac{i}{\hbar}\int_0^t  p(\tau) \dot{x}(\tau) -\frac{p(\tau)^2}{2m} d\tau +\frac{1}{2}\int_0^t \dot{x}(\tau)^2 +p(\tau)^2 d\tau\right)\\ \nonumber
\cdot \exp\left(-\frac{i}{\hbar} \int_0^t V(x(\tau),p(\tau),\tau) \, d\tau\right) \cdot \delta(x(t)-y)
\end{align}
as a Hida distribution. In this expression the sum of the first and the third integral in the exponential is the action $S(x,p)$, and the delta function (Donsker's delta) serves to pin trajectories to $y$ at time $t$. The second integral is introduced to simulate the Lebesgue integral by a local compensation of the fall-off of the Gaussian reference measure $\mu$.
Furthermore we use a Brownian motion starting in $y_0$ as position variable and the momentum variable is modeled by white noise, i.e.~
\begin{eqnarray}\label{varchoice}
x(\tau)=y(0)+\sqrt{\frac{\hbar}{m}}B(\tau),\quad 
p(\tau) =\omega(\tau),\quad 0\leq \tau \leq t. 
\end{eqnarray}
The construction is done in terms of the $T$-transform (infinite-dimensional version of the Fourier transform w.r.t~a Gaussian measure), which characterizes Hida distributions, see Theorem \ref{charthm}. At the same time, the $T$-transform of the constructed Feynman integrands provides us with their generating functional. Finally using the generating functional, we can show that the generalized expectation (generating functional at zero) gives the Greens function to the corresponding Schrödinger equation. Moreover, with help of the generating functional we can show that the canonical commutation relations are fulfilled. This confirms on a mathematical rigorous level the heuristics developed in \cite{FeHi65}.

These are the core results of this article:
\begin{itemize}
\item The concepts of generalized Gauss kernels from \cite{GS98a} are extended to the vector--valued case explicitly.
\item The concepts for realizing the Feynman integrands in phase space in White Noise Analysis are provided.
\item The free Feynman integrand $I_0$ and the Feynman integrand for the harmonic oscillator $I_{HO}$ in phase space are constructed as Hida distributions, see Theorem \ref{freethm} and Theorem \ref{hothm}.
\item The results in Theorem \ref{freethm} and Theorem \ref{hothm} provide us with the generating functional to the Feynman integrands. The generalized expectations (generating functional at zero) provide us the Greens functions to the corresponding Schrödinger equation.
\item In Theorem \ref{commthm} the canonical commutator relations for $I_0$ are obtained in the sense of Feynman and Hibbs, see \cite{FeHi65}.
\end{itemize}

\section{White Noise Analysis}
\subsection{Gel'fand Triples}
Starting point is the Gel'fand triple $S_d(\R) \subset L^2_d(\R,dx) \subset S'_d(\R)$ of the $\R^d$-valued, Schwartz test functions and tempered distributions with the Hilbert space of (equivalence classes of) $\R^d$-valued square integrable functions w.r.t.~the Lebesgue measure as central space (equipped with its canonical inner product $(\cdot, \cdot)$ and norm $\|\cdot\|$) , more detailed see e.g.~ \cite[Exam.~11]{W95}.
Since $S_d(\R)$ is a nuclear space, represented as projective limit of a decreasing chain of Hilbert spaces $(H_p)_{p\in \N}$, see e.g.~\cite[Chap.~2]{RS75a} and \cite{GV68} i.e.~
\begin{equation*}
S_d(\R) = \bigcap_{p \in \N} H_p,
\end{equation*}
we have that $S_d(\R)$ is a countably Hilbert space in the sense of Gel'fand and Vilenkin \cite{GV68}. We denote the inner product and the corresponding norm on $H_p$ by $(\cdot,\cdot)_p$ and $\|\cdot\|_p$, respectively, with the convention $H_0 = L^2(\R, dx)$.
Let $H_{-p}$ be the dual space of $H_p$ and let $\langle \cdot , \cdot \rangle$ denote the dual pairing on $H_{p} \times H_{-p}$. $H_{p}$ is continuously embedded into $L^2_d(\R,dx)$. By identifying $L_d^2(\R,dx)$ with its dual $L_d^2(\R,dx)'$, via the Riesz isomorphism, we obtain the chain $H_p \subset L_d^2(\R, dx) \subset H_{-p}$.
Note that $\displaystyle S'_d(\R)= \bigcup_{p\in \N} H_{-p}$, i.e.~$S'_d(\R)$ is the inductive limit of the increasing chain of Hilbert spaces $(H_{-p})_{p\in \N}$, see  e.g.~\cite{GV68}.
We denote the dual pairing of $S_d(\R)$ and $S'_d(\R)$ also by $\langle \cdot , \cdot \rangle$. Note that its restriction on $S_d(\R) \times L_d^2(\R, dx)$ is given by $(\cdot, \cdot )$.
We also use the complexifications of these spaces denoted with the subindex $\C$ (as well as their inner products and norms). The dual pairing we extend in a bilinear way. Hence we have the relation 
\begin{equation*}
\langle g,f \rangle = (g,\overline{f}), \quad f,g \in L^2(\R)_{\C},
\end{equation*}
where the overline denotes the complex conjugation.
\subsection{White Noise Spaces}
We consider on $S_d' (\R)$ the $\sigma$-algebra $\cC_{\sigma}(S_d' (\R))$ generated by the cylinder sets $\{ \omega \in S_d' (\R) | \langle \xi_1, \omega \rangle \in F_1, \dots ,\langle \xi_n, \omega \rangle \in F_n\} $, $\xi_i \in S_d(\R)$, $ F_i \in \cB(\R),\, 1\leq i \leq n,\, n\in \N$, where $\cB(\R)$ denotes the Borel $\sigma$-algebra on $\R$.\\
\noindent The canonical Gaussian measure $\mu$ on $C_{\sigma}(S_d'(\R))$ is given via its characteristic function
\begin{eqnarray*}
\int_{S_d' (\R)} \exp(i \langle {\bf f}, \omega \rangle ) d\mu(\omega) = \exp(- \tfrac{1}{2} \| {\bf f}\|^2 ), \;\;\; {\bf f} \in S_d(\R),
\end{eqnarray*}
\noindent by the theorem of Bochner and Minlos, see e.g.~\cite{Mi63}, \cite[Chap.~2 Theo.~1.~11]{BK95}, and \cite{ HKPS93}. The space $(S_d'(\R),\cC_{\sigma}(S_d'(\R)), \mu)$ is the ba\-sic probability space in our setup.
The cen\-tral Gaussian spa\-ces in our frame\-work are the Hil\-bert spaces $(L^2):= L^2(S_d'(\R),$ $\cC_{\sigma}(S_d' (\R)),\mu)$ of complex-valued square in\-te\-grable func\-tions w.r.t.~the Gaussian measure $\mu$.\\
Within this formalism a version of a d-dimensional Brownian motion is given by 
\begin{equation}\label{BrownianMotion}
{\bf B}(t,{\boldsymbol \omega}) := ( \langle  \1_{[0,t)},\omega_1 \rangle, \dots  \langle  \1_{[0,t)},\omega_d \rangle), \quad {\boldsymbol \omega}=(\omega_1,\dots \omega_d) \in S'_d(\R),\quad t \geq 0,
\end{equation}
in the sense of an $(L^2)$-limit. Here $\1_A$ denotes the indicator function of a set $A$. 

\subsection{The Hida triple}

Let us now consider the complex Hilbert space $(L^2)$ and the corresponding Gel'fand triple
\begin{equation*}
(S) \subset (L^2) \subset (S)'.
\end{equation*}
Here $(S)$ denotes the space of Hida test functions and $(S)'$ the space of Hida distributions. In the following we denote the dual pairing between elements of $(S)$ and $(S)'$ by $\langle \! \langle \cdot , \cdot \rangle \!\rangle$. 
Instead of reproducing the construction of $(S)'$ here we give its characterization in terms of the $T$-transform.\\
\begin{definition}
We define the $T$-transform of $\Phi \in (S)'$ by
\begin{equation*}
T\Phi({\bf f}) := \langle\!\langle \Phi, \exp(i \langle {\bf f}, \cdot \rangle) \rangle\!\rangle, \quad  {\bf f}:= ({ f_1}, \dots ,{ f_d }) \in S_{d}(\R).
\end{equation*}
\end{definition}

\begin{remark}
\begin{itemize}
\item[(i)] Since $\exp(i \langle {\bf f},\cdot \rangle) \in (S)$ for all $f \in S_d(\R)$, the $T$-transform of a Hida distribution is well-defined.
\item[(ii)] For ${\bf f} = 0$ the above expression yields $\langle\!\langle \Phi, 1 \rangle\!\rangle$, therefore $T\Phi(0)$ is called the generalized expectation of $\Phi \in (S)'$.
\end{itemize}
\end{remark}

\noindent In order to characterize the space $(S)'$ by the $T$-transform we need the following definition.

\begin{definition}
A mapping $F:S_{d}(\R) \to \C$ is called a {\emph U-functional} if it satisfies the following conditions:
\begin{itemize}
\item[U1.] For all ${\bf{f, g}} \in S_{d}(\R)$ the mapping $\R \ni \lambda \mapsto F(\lambda {\bf f} +{\bf g} ) \in \C$ has an analytic continuation to $\lambda \in \C$ ({\bf{ray analyticity}}).
\item[U2.] There exist constants $0<K,C<\infty$ and a $p \in \N_0$ such that 
\begin{equation*}
|F(z{\bf f})|\leq K\exp(C|z|^2 \|{\bf f} \|_p^2), 
\end{equation*}
for all $z \in \C$ and ${\bf f} \in S_{d}(\R)$ ({\bf{growth condition}}).
\end{itemize}
\end{definition}

\noindent This is the basis of the following characterization theorem. For the proof we refer to \cite{PS91,Kon80,HKPS93,KLPSW96}.

\begin{theorem}\label{charthm}
A mapping $F:S_{d}(\R) \to \C$ is the $T$-transform of an element in $(S)'$ if and only if it is a U-functional.
\end{theorem}
Theorem \ref{charthm} enables us to discuss convergence of sequences of Hida distributions by considering the corresponding $T$-transforms, i.e.~ by considering convergence on the level of U-functionals. The following corollary is proved in \cite{PS91,HKPS93,KLPSW96}.

\begin{corollary}\label{seqcor}
Let $(\Phi_n)_{n\in \N}$ denote a sequence in $(S)'$ such that
\begin{itemize}
\item[(i)] For all ${\bf f} \in S_{d}(\R)$, $((T\Phi_n)({\bf f}))_{n\in \N}$ is a Cauchy sequence in $\C$.
\item[(ii)] There exist constants $0<C,D<\infty$ such that for some $p \in \N_0$ one has 
\begin{equation*}
|(T\Phi_n)(z{\bf f })|\leq D\exp(C|z|^2\|{\bf f}\|_p^2)
\end{equation*}
for all ${\bf f} \in S_{d}(\R),\, z \in \C$, $n \in \N$.
\end{itemize}
Then $(\Phi_n)_{n\in \N}$ converges strongly in $(S)'$ to a unique Hida distribution.
\end{corollary}

\begin{example}[Vector valued white noise]
\noindent Let $\,{\bf{B}}(t)$, $t\geq 0$ be the $d$-di\-men\-sional Brow\-nian motion as in \eqref{BrownianMotion}. 
Consider $\frac{{\bf{B}}(t+h,\boldsymbol{\omega}) - {\bf{B}}(t,\boldsymbol{\omega})}{h} = (\langle \frac{\1_{[t,t+h)}}{h} , \omega_1 \rangle , \dots (\langle \frac{\1_{[t,t+h)}}{h} , \omega_d \rangle)$, $h>0$. 
Then in the sense of Corollary \ref{seqcor} it exists
\begin{eqnarray*}
\langle \delta_t, {\boldsymbol \omega} \rangle := (\langle \delta_t,\omega_1 \rangle, \dots ,\langle \delta_t,\omega_d \rangle):= \lim_{h\searrow 0} \frac{{\bf{B}}(t+h,\boldsymbol{\omega}) - {\bf{B}}(t,\boldsymbol{\omega})}{h}.
\end{eqnarray*}
Of course for the left derivative we get the same limit. Hence it is natural to call the generalized process $\langle \delta_t, {\boldsymbol \omega} \rangle$, $t\geq0$ in $(S)'$ vector valued white noise. One also uses the notation ${\boldsymbol \omega}(t) =\langle \delta_t, {\boldsymbol \omega} \rangle$, $t\geq 0$. 
\end{example}

Another useful corollary of Theorem \ref{charthm} concerns integration of a family of generalized functions, see \cite{PS91,HKPS93,KLPSW96}.
\begin{corollary}\label{intcor}
Let $(\Lambda, \cA, \nu)$ be a measure space and $\Lambda \ni\lambda \mapsto \Phi(\lambda) \in (S)'$ a mapping. We assume that its $T$--transform $T \Phi$ satisfies the following conditions:
\begin{enumerate}
\item[(i)] The mapping $\lambda \mapsto T(\Phi(\lambda))({\bf f})$ is measurable for all ${\bf f} \in S_d(\R)$.
\item[(ii)] There exists a $p \in \N_0$ and functions $C \in L^{\infty}(\Lambda, \nu)$ and $D \in L^1(\Lambda, \nu)$ such that 
\begin{equation*}
   \abs{T(\Phi(\lambda))(z{\bf f})} \leq D(\lambda)\exp(C(\lambda) \abs{z}^2 \norm{{\bf f}}^2), 
\end{equation*}
for a.e.~$ \lambda \in \Lambda$ and for all ${\bf f} \in S_d(\R)$, $z\in \C$.
\end{enumerate}
Then, in the sense of Bochner integration in $H_{-q} \subset (S)'$ for a suitable $q\in \N_0$, the integral of the family of Hida distributions is itself a Hida distribution, i.e.~$\!\displaystyle \int_{\Lambda} \Phi(\lambda) \, d\nu(\lambda) \in (S)'$ and the $T$--transform interchanges with integration, i.e.~
\begin{equation*}
   T\lb \int_{\Lambda} \Phi(\lambda) \, d\nu(\lambda) \rb =
   	\int_{\Lambda} T(\Phi(\lambda)) \, d\nu(\lambda).
\end{equation*}
\end{corollary}

Based on the above theorem, we introduce the following Hida distribution.
\begin{definition}
\label{D:Donsker} 
We define Donsker's delta at $x \in \R$ corresponding to $0 \neq {\boldsymbol\eta} \in L_{d}^2(\R)$ by
\begin{equation*}
   \delta_x(\lab {\boldsymbol\eta},\cdot \rab) := 
   	\frac{1}{2\pi} \int_{\R} \exp(i \lambda (\lab {\boldsymbol\eta},\cdot \rab -x)) \, d \lambda
\end{equation*}
in the sense of Bochner integration, see e.g.~\cite{HKPS93,LLSW94,W95}. Its $T$--transform in ${\bf f} \in S_d(\R)$ is given by
\begin{equation*}
   T(\delta_x(\langle  {\boldsymbol\eta},\cdot \rangle)({\bf f}) 
   	= \frac{1}{\sqrt{2\pi \langle {\boldsymbol\eta}, {\boldsymbol\eta}\rangle}} \exp\left( -\frac{1}{2\langle {\boldsymbol\eta},{\boldsymbol\eta} \rangle}(i\lab {\boldsymbol\eta},{\bf f} \rab - x)^2 -\frac{1}{2}\langle {\bf f},{\bf f}\rangle \right)\;.
\end{equation*}
\end{definition}

\subsection{Generalized Gauss Kernels}
Here we review a special class of Hida distributions which are defined by their T-transform, see e.g.~\cite{GS98a}. Let $\mathcal{B}$ be the set of all continuous bilinear mappings $B:S_{d}(\R) \times S_{d}(\R) \to \C$. Then the functions
\begin{equation*}
S_d(\R)\ni f \mapsto \exp\left(-\frac{1}{2} B({\bf f},{\bf f})\right) \in \C
\end{equation*}
for all $B\in \mathcal{B}$ are U-functionals. Therefore, by using the characterization of Hida distributions in Theorem \ref{charthm},
the inverse T-transform of these functions 
\begin{equation*}
\Phi_B:=T^{-1} \exp\left(-\frac{1}{2} B\right)
\end{equation*}
are elements of $(S)'$.

\begin{definition}\label{GGK}
The set of {\bf{generalized Gauss kernels}} is defined by
\begin{equation*}
GGK:= \{ \Phi_B,\; B\in \mathcal{B} \}.
\end{equation*}
\end{definition}

\begin{example}{\cite{GS98a}} \label{Grotex} We consider a symmetric trace class operator $K$ on $L^2_{d}(\R)$ such that $-\frac{1}{2}<K\leq 0$, then
\begin{align*}
\int_{S'_{d}(\R)} \exp\left(- \langle \omega,K \omega\rangle \right) \, d\mu(\omega) 
= \left( \det(Id +2K)\right)^{-\frac{1}{2}} < \infty.
\end{align*}
For the definition of $\langle \cdot,K \cdot \rangle$ see the remark below.
Here $Id$ denotes the identity operator on the Hilbert space $L^2_{d}(\R)$, and $\det(A)$ of a symmetric trace class operator $A$ on $L^2_{d}(\R)$ denotes the infinite product of its eigenvalues, if it exists. In the present situation we have $\det(Id +2K)\neq 0$.
There\-fore we obtain that the exponential $g= \exp(-\frac{1}{2} \langle \cdot,K \cdot \rangle)$ is square-integrable and its T-transform is given by 
\begin{equation*}
Tg({\bf f}) = \left( \det(Id+K) \right)^{-\frac{1}{2}} \exp\left(-\frac{1}{2} ({\bf f}, (Id+K)^{-1} {\bf f})\right), \quad {\bf f} \in S_{d}(\R).
\end{equation*}
Therefore $\left( \det(Id+K) \right)^{\frac{1}{2}}g$ is a generalized Gauss kernel.
\end{example}

\begin{remark}
Since a trace class operator is compact, see e.g.~\cite{RS75a}, we have that $K$ in the above example is diagonalizable, i.e.~
\begin{equation*}
Kf = \sum_{k=1}^{\infty} k_n (f,e_n)e_n, \quad f \in L^2(\R,dx),
\end{equation*}
where $(e_n)_{n\in \N}$ denotes an eigenbasis of the corresponding eigenvalues $(k_n)_{n\in \N}$ with $k_n \in (-\frac{1}{2}, 0 ]$, for all $n \in \N$. Since $K$ is compact, we have that $\lim\limits_{n\to \infty} k_n =0$ and since $K$ is trace class we also have $\sum_{n=1}^{\infty} (e_n, -K e_n)< \infty$. We define for $\omega \in S_d'(\R)$
\begin{eqnarray*} 
- \langle \omega, K \omega \rangle := \lim_{N \to \infty} \sum_{n=1}^N \langle e_n, \omega\rangle (-k_n)\langle e_n,\omega \rangle. 
\end{eqnarray*}
Then as a limit of measurable functions $\omega \mapsto -\langle \omega, K \omega \rangle$  is measurable and hence 
\begin{eqnarray*} 
\int\limits_{S_d'(\R)} \exp(-  \langle \omega, K \omega \rangle ) \, d\mu(\omega) \in [0, \infty].
\end{eqnarray*}
The explicit formula for the $T$-transform and expectation then follow by a straightforward calculation with help of the above limit procedure. 
\end{remark}

\begin{definition}
Let $K: L^2_{d,\C}(\R, dx) \to L^2_{d,\C}(\R, dx)$ be linear and continuous such that
\begin{itemize}
\item[(i)] $Id+K$ is injective, 
\item[(ii)] there exists $p \in \N_0$ such that $(Id+K)(L^2_{d,\C}(\R,\,dx)) \subset H_{p,\C}$ is dense,
\item[(iii)] there exist $q \in\N_0$ such that $(Id+K)^{-1} :H_{p,\C} \to H_{-q,\C}$ is continuous with $p$ as in (ii).
\end{itemize}
Then we define the normalized exponential
\begin{equation}\label{Nexp}
{\rm{Nexp}}(- \frac{1}{2} \langle \cdot ,K \cdot \rangle)
\end{equation}
by
\begin{align*}
T({\rm{Nexp}}(- \frac{1}{2} \langle \cdot ,K \cdot \rangle))({\bf f}) &:= \exp(-\frac{1}{2} \langle {\bf f}, (Id+K)^{-1} {\bf f} \rangle),\quad {\bf f} \in S_d(\R).
\end{align*}
\end{definition}

\begin{remark}
The "normalization" of the exponential in the above definition can be regarded as a division of a divergent factor. In an informal way one can write
\begin{align*}
T({\rm{Nexp}}(- \frac{1}{2} \langle \cdot ,K \cdot \rangle))({\bf f})=\frac{T(\exp(- \frac{1}{2} \langle \cdot ,K \cdot \rangle))(f)}{T(\exp(- \frac{1}{2} \langle \cdot ,K \cdot \rangle))(0)}=\frac{T(\exp(- \frac{1}{2} \langle \cdot ,K \cdot \rangle))(f)}{\sqrt{\det(Id+K)}} , \quad {\bf f} \in S_d(\R), 
\end{align*}
i.e.~ if the determinant in the Example \ref{Grotex} above is not defined, we can still define the normalized exponential by the T-transform without the diverging prefactor. The assumptions in the above definition then guarantee the existence of the generalized Gauss kernel in \eqref{Nexp}.
\end{remark}

\begin{example}\label{pointprod}
	For sufficiently "nice" operators $K$ and $L$ on $L^2_{d}(\R)_{\C}$ we can define the product 
			\begin{equation*}
				{\rm{Nexp}}\big( - \frac{1}{2} \langle \cdot,K \cdot \rangle  \big) \cdot \exp\big(-\frac{1}{2} \langle \cdot,L\cdot \rangle \big)
			\end{equation*}
	of two square-integrable functions. Its $T$-transform is then given by 
			\begin{multline*}
				T\Big({\rm{Nexp}}( - \frac{1}{2} \langle \cdot,K \cdot\rangle ) \cdot \exp( - \frac{1}{2} \langle \cdot,L \cdot\rangle )\Big)({\bf f})\\
				=\sqrt{\frac{1}{\det(Id+L(Id+K)^{-1})}}
				\exp(-\frac{1}{2} \langle {\bf f}, (Id+K+L)^{-1} {\bf f} \rangle ),\quad {\bf f} \in S_{d}(\R),
			\end{multline*}	
	in the case the left hand side indeed is a U-funcional.		
\end{example}

In the case $g \in S_d(\R)$, $c\in\C$ the product between the Hida distribution $\Phi$ and the Hida test function $\exp(i \langle g,. \rangle + c)$ is well defined because $(S)$ is a continuous algebra under pointwise multiplication.  The next definition is an extension of this product.

\begin{definition}\label{linexp}
The pointwise product of a Hida distribution $\Phi \in (S)'$ with an exponential of a linear term, i.e.~
\begin{equation*}
\Phi \cdot \exp(i \langle {\bf g}, \cdot \rangle  +c), \quad {\bf g} \in L^2_{d}(\R)_{\C}, \, c \in \C,
\end{equation*}
is defined by 
\begin{equation*}
T(\Phi \cdot \exp(i\langle  {\bf g}, \cdot \rangle  + c))({\bf f}):= T\Phi({\bf f}+{\bf g})\exp(c),\quad {\bf f} \in S_d(\R),  
\end{equation*}
if $T\Phi$ has a continuous extension to $L^2_d(\R)_{\C}$ and the term on the right-hand side is a U-functional in ${\bf f} \in S_d(\R)$.
\end{definition}

\begin{definition}\label{donsker}
Let $D \subset \R$ such, that $0 \in \overline{D}$. Under the assumption that $T\Phi$ has a continuous extension to $L^2_d(\R)_{\C}$, ${\boldsymbol\eta}\in L^2_d(\R)_{\C}$, $y \in \R$, $\lambda \in \gamma_{\alpha}:=\{\exp(-i\alpha)s|\, s \in \R\}$ and that the integrand 
\begin{equation*}
\gamma_{\alpha} \ni \lambda \mapsto \exp(-i\lambda y)T\Phi({\bf f}+\lambda {\boldsymbol\eta}) \in \C
\end{equation*}
fulfills the conditions of Corollary \ref{intcor} for all $\alpha \in D$, one can define the product 
\begin{equation*}
\Phi \cdot \delta(\langle {\boldsymbol\eta}, \cdot \rangle-y),
\end{equation*}
by
\begin{equation*}
T(\Phi \cdot \delta(\langle {\boldsymbol\eta}, \cdot \rangle-y))({\bf f})
:= \lim_{\alpha \to 0} \int_{\gamma_{\alpha}} \exp(-i \lambda y) T\Phi({\bf f}+\lambda {\boldsymbol\eta}) \, d \lambda.
\end{equation*}
Of course under the assumption that the right-hand side converges in the sense of Corollary \ref{seqcor}, see e.g.~\cite{GS98a}.
\end{definition}

\begin{lemma}\label{thelemma}
Let  $L$ be a $d\times d$ block operator matrix on $L^2_{d}(\R)_{\C}$ acting componentwise such that all entries are bounded operators on $L^2(\R)_{\C}$.
Let $K$ be a d $\times d$ block operator matrix on $L^2_{d}(\R)_{\C}$, such that $Id+K$ and $N=Id+K+L$ are bounded with bounded inverse. Furthermore assume that $\det(Id+L(Id+K)^{-1})$ exists and is different from zero (this is e.g.~the case if $L$ is trace class and -1 in the resolvent set of $L(Id+K)^{-1}$).
Let $M_{N^{-1}}$ be the matrix given by an orthogonal system $({\boldsymbol\eta}_k)_{k=1,\dots J}$ of non--zero functions from $L^2_d(\R)$, $J\in \N$, under the bilinear form $\left( \cdot ,N^{-1} \cdot \right)$, i.e.~ $(M_{N^{-1}})_{i,j} = \left( {\boldsymbol\eta}_i ,N^{-1} {\boldsymbol\eta}_j \right)$.
Under the assumption that either 
\begin{eqnarray*}
\Re(M_{N^{-1}}) >0 \quad \text{ or }\quad \Re(M_{N^{-1}})=0 \,\text{ and } \,\Im(M_{N^{-1}}) \neq 0,
\end{eqnarray*} 
where $M_{N^{-1}}=\Re(M_{N^{-1}}) + i \Im(M_{N^{-1}})$ with real matrices $\Re(M_{N^{-1}})$ and $\Im(M_{N^{-1}})$, \\
then
\begin{eqnarray*}
\Phi_{K,L}:={\rm Nexp}\big(-\frac{1}{2} \langle \cdot, K \cdot \rangle \big) \cdot \exp\big(-\frac{1}{2} \langle \cdot, L \cdot \rangle \big) \cdot \exp(i \langle \cdot, {\bf g} \rangle)
\cdot \prod_{i=1}^J \delta (\langle \cdot, {\boldsymbol\eta}_k \rangle-y_k),
\end{eqnarray*}
for ${\bf g} \in L^2_{d}(\R,\C),\, t>0,\, y_k \in \R,\, k =1\dots,J$, exists as a Hida distribution. \\
Moreover for ${\bf f} \in S_d(\R)$
\begin{multline*}
T\Phi_{K,L}({\bf f})=\frac{1}{\sqrt{(2\pi)^J  \det((M_{N^{-1}}))}}
\sqrt{\frac{1}{\det(Id+L(Id+K)^{-1})}}\\ 
\times \exp\bigg(-\frac{1}{2} \big(({\bf f}+{\bf g}), N^{-1} ({\bf f}+{\bf g})\big) \bigg)
\exp\bigg(-\frac{1}{2} (u,(M_{N^{-1}})^{-1} u)\bigg),
\end{multline*}
where
\begin{equation*}
u= \left( \big(iy_1 +({\boldsymbol\eta}_1,N^{-1}({\bf f}+{\bf g})) \big), \dots, \big(iy_J +({\boldsymbol\eta}_J,N^{-1}({\bf f}+{\bf g})) \big) \right)
\end{equation*}
\end{lemma}

\noindent{{\bf{Proof:}}
We want to give meaning to the expression
\begin{equation*}
{\rm Nexp}\big(-\frac{1}{2} \langle \cdot, K \cdot \rangle \big) \cdot \exp\big(-\frac{1}{2} \langle \cdot, L \cdot \rangle \big) \cdot \exp(i \langle \cdot, {\bf g} \rangle)
\cdot \prod_{k=1}^J \delta (\langle \cdot, {\boldsymbol\eta}_k \rangle-y_k),
\end{equation*}
\vspace*{-1pt}
\noindent using Definition \ref{donsker} inductively. 
Note that ${\rm Nexp}\big(-\frac{1}{2} \langle \cdot, K \cdot \rangle \big) \cdot \exp\big(-\frac{1}{2} \langle \cdot, L \cdot \rangle \big)$ can be defined as in Example \ref{pointprod}.
Hence we obtain for the T-transform of the integrand 
\begin{multline*}
\gamma_{\alpha}^J\ni \lambda \mapsto \Phi_{\lambda} =\exp(-i \sum_{j=1}^J \lambda_j e^{-i\alpha} y_j) \cdot \exp(i \sum_{j=1}^J \lambda_j e^{-i\alpha} \langle {\boldsymbol\eta}_j, \cdot \rangle)\\
\cdot {\rm{N}}\exp(-\frac{1}{2} \langle \cdot ,K \cdot \rangle ) \exp(-\frac{1}{2} \langle \cdot ,L \cdot \rangle )\exp(i \langle \cdot, {\bf g} \rangle) 
\end{multline*}
in ${\bf f} \in S_{d}(\R)$,
\begin{multline*}
T\big(\exp(-i \sum_{j=1}^J \lambda_j e^{-i\alpha} y_j) \exp(i \sum_{j=1}^J \lambda_j e^{-i\alpha} \langle {\boldsymbol\eta}_j, \cdot \rangle)\\
\cdot {\rm{N}}\exp(-\frac{1}{2} \langle \cdot ,K \cdot \rangle )\cdot \exp(-\frac{1}{2} \langle \cdot ,L \cdot \rangle )\cdot\exp(i \langle \cdot, {\bf g} \rangle)\big) ({\bf f})\\
= \exp(-i \sum_{j=1}^J \lambda_j e^{-i\alpha} y_j) \\
\times T\big({\rm{N}}\exp(-\frac{1}{2} \langle \cdot ,K \cdot \rangle )\cdot \exp(-\frac{1}{2} \langle \cdot ,L \cdot \rangle )\cdot\exp(i \langle \cdot, {\bf g}+\sum_{j=1}^J \lambda_j e^{-i\alpha} \langle {\boldsymbol\eta}_j \rangle)\big)({\bf f})\\
=  \exp(-i \sum_{j=1}^J \lambda_j e^{-i\alpha} y_j) \frac{1}{\sqrt{ \det( Id+L(Id+K)^{-1})}} \\
\times \exp(-\frac{1}{2} \big(({\bf f}+ {\bf g}+\sum_{j=1}^J \lambda_j e^{-i\alpha} {\boldsymbol\eta}_j), N^{-1} ({\bf f}+ {\bf g}+\sum_{k=1}^J \lambda_k e^{-i\alpha} {\boldsymbol\eta}_k)\!\big) 
\end{multline*}
Here we use $y = (y_1, \dots y_J)$ and $\lambda =(\lambda_1, \dots \lambda_J)$, respectively. Then we can rewrite the above formula with the help of the matrix $M_{N^{-1}}$ as
\begin{multline}\label{Integranddons}
\gamma_{\alpha}^J\ni \lambda \mapsto T\big(\exp(-i \sum_{j=1}^J \lambda_j e^{-i\alpha} y_j) \exp(i \sum_{j=1}^J \lambda_j e^{-i\alpha} \langle {\boldsymbol\eta}_j, \cdot \rangle)\\
\cdot {\rm{N}}\exp(-\frac{1}{2} \langle \cdot,K \cdot \rangle ) \cdot\exp(-\frac{1}{2} \langle \cdot ,L \cdot \rangle )\cdot\exp(i \langle \cdot, {\bf g} \rangle)\big) ({\bf f})\\
= \frac{1}{\sqrt{ \det( Id+L(Id+K)^{-1})}}  \exp\Bigg(-\frac{1}{2} e^{-2i\alpha} (\lambda,M_{N^{-1}} \lambda) \\
-e^{-i\alpha} \lambda\bigg(\Big( \big(({\bf f}+ {\bf g}),N^{-1} {\boldsymbol\eta}_1\big), \dots \big(({\bf f}+ {\bf g}),N^{-1} {\boldsymbol\eta}_d\big)
\Big)-iy\bigg)\Bigg).
\end{multline}
The function in \eqref{Integranddons} is integrable w.r.t.~the Lebesgue measure, if the real part of $e^{-2i\alpha} M_{N^{-1}}$, i.e.~$\Re(e^{-2i\alpha} M_{N^{-1}}) =\cos(2\alpha) \Re(M_{N^{-1}}) + \sin(2\alpha)\Im(M_{N^{-1}})$, is positive definite. Our assumptions on $M_{N^{-1}}$ in Lemma \ref{thelemma} imply that this holds for $\alpha$ in a set $D$, as required in Definition \ref{donsker}.  The calculation of the T-transform then follows in analogous way to the calculation of the T-transform of a product of Donskers delta functions, see e.g.~\cite{LLSW94, W95}.
\hfill $\blacksquare$

\section{Phase space Feynman path integrals}
In the following we realize rigorously the ansatz 
\begin{align}\label{anpsfey}
I_V = {\rm Nexp}\left( \frac{i}{\hbar}\int_0^t  p(\tau) \dot{x}(\tau) -\frac{p(\tau)^2}{2m} d\tau +\frac{1}{2}\int_0^t \dot{x}(\tau)^2 +p(\tau)^2 d\tau\right)\\ \nonumber
\times \exp\left(-\frac{i}{\hbar} \int_0^t V(x(\tau),p(\tau),\tau) \, d\tau\right) \cdot \delta(x(t)-y),
\end{align}
for the Feynman integrand in phase space for $V=0$ (free particle) and the harmonic oscillator, i.e.~$x\mapsto V(x)=\tfrac{1}{2}k x^2,\, k\geq 0$, motivated in the introduction, see \eqref{psfey}. 
\subsection{The free Feynman integrand in phase space}

First we consider $V=0$ (free particle). For simplicity let $\hbar=m=1$ and $y_0=0$. Furthermore we choose to have one space dimension and one dimension for the corresponding momentum variable, i.e.~the underlying space is $S_2(\R)$. Note that the first term in \eqref{anpsfey} can be considered as a exponential of a quadratic type:
\begin{multline*}
{\rm Nexp}\left( i\int_0^t  (p(\tau) \dot{x}(\tau) -\frac{p(\tau)^2}{2}) d\tau +\frac{1}{2}\int_0^t \dot{x}(\tau)^2 +p(\tau)^2 d\tau\right)\\
={\rm Nexp}\bigg(-\frac{1}{2} \big\langle
(\omega_x,\omega_p),K(
    \omega_x,
    \omega_p )\big\rangle\bigg),
\end{multline*}
where the operator matrix $K$ on $L_2^2(\R)_{\C}$ can be written as
\begin{equation}\label{kinmat}
K=\left(
\begin{array}[h]{l l}
    -\1_{[0,t)}&-i \1_{[0,t)}\\[0,1 cm]
    -i \1_{[0,t)}& -(1-i) \1_{[0,t)}
\end{array}
\right).
\end{equation}
Here the operator $\1_{[0,t)}$ denotes the multiplication with $\1_{[0,t)}$.
Hence, the integrand in \eqref{anpsfey} can then be written as
\begin{equation*}
I_0={\rm{Nexp}}\left(-\frac{1}{2} \langle (\omega_x,\omega_p), K (\omega_x,\omega_p) \rangle\right) \cdot \delta \bigg( \langle (\omega_x, \omega_p), (\1_{[0,t)}, 0) \rangle-y\bigg),
\end{equation*}
where the last term pins the position variable to $y$ at $t$. Note that the momentum variable is not pinned. Our aim is to apply Lemma \ref{thelemma} with $K$ as above and ${\bf g}=0$, $L=0$ and as ${\boldsymbol\eta}= (\1_{[0,t)} ,0)$.
The inverse of $(Id+K)$ is given by
\begin{eqnarray}
N^{-1}=(Id+K)^{-1}=\label{InvN}
\bigg(
\begin{array}{l l}
 \1_{[0,t)^c}+i \1_{[0,t)}&i \1_{[0,t)}\\
    i \1_{[0,t)}& \1_{[0,t)^c}
\end{array} 
\bigg),
\end{eqnarray}
hence $({\boldsymbol\eta},N^{-1}{\boldsymbol\eta})=i\cdot t$. Therefore the assumptions of Lemma \ref{thelemma} are fulfilled. Thus $I_0$ exists as a Hida distribution. By applying Lemma \ref{thelemma} its $T$-transform in $(f_x,f_p) \in S_2(\R)$ is given by 
\begin{align}
&T\left({\rm{Nexp}}\left(-\frac{1}{2} \langle (\omega_x,\omega_p), K (\omega_x,\omega_p) \rangle\right) \cdot \delta \left( \langle (\omega_x, \omega_p), (\1_{[0,t{)}}, 0) \rangle-y\right)\right)(f_x,f_p) \nonumber\\ 
&=\nonumber \frac{1}{\sqrt{2 \pi i t}} \exp \left( -\frac{1}{2it}\left(y-\int_0^t f_x+f_p\,ds\right)^2- \frac{1}{2} \left( (f_x,f_p), N^{-1} (f_x,f_p) \right)\right)\\\nonumber
&=\nonumber\frac{1}{\sqrt{2 \pi i t}} \exp \left( -\frac{1}{2it}\left(y-\int_0^t f_x+f_p\,ds\right)^2\right) \\
&\hspace{1 cm}\times \nonumber\exp \left(
- \frac{1}{2} \left( (f_x,f_p), \bigg(
\begin{array}{l l}
 \1_{[0,t)^c}+i \1_{[0,t)}&i \1_{[0,t)}\\
    i \1_{[0,t)}& \1_{[0,t)^c}
\end{array} 
\bigg)(f_x,f_p) \right)\right)\\\nonumber
&=\frac{1}{\sqrt{2 \pi i t}} \exp \left( -\frac{1}{2it}\left(y-\int_0^t f_x+f_p\,ds\right)^2\right)\\
&\hspace{1 cm}\times \exp\left(-\frac{1}{2} \left( \int_{[0,t)^c} f_x^2+f_p^2\, ds +i \int_{[0,t)} f_x^2 \, ds +2i \int_{[0,t)} f_x(s)f_p(s) \, ds \right) \right)\label{genfun}.
\end{align}
Hence its generalized expectation %\ref{genexpec}
\begin{equation*}
\mathbb{E}(I_0) = TI_0(0) = \frac{1}{\sqrt{2 \pi i t}} \exp(-\frac{1}{2it}y^2)=K(y,t,0,0)
\end{equation*}
gives indeed the Greens function to the Schrödinger equation for a free particle. Summarizing we have the following Theorem:

\begin{theorem}\label{freethm}
Let $y\in \R$, $0<t<\infty$, then the free Feynman integrand in phase space $I_0$ exists as a Hida distribution. Its generating functional $TI_0$ is given by \eqref{genfun} and its generalized expectation $\mathbb{E}(I_0)=TI_0(0)$ is the Greens function to the Schrödinger equation for the free particle.
\end{theorem}

\subsection{The Feynman-integrand for the harmonic oscillator in phase space}
In this section we construct the Feynman integrand for the harmonic oscillator in phase space. I.e.~the potential is given by $x \mapsto V(x)= \frac{1}{2}k x^2$, $0 \leq k<\infty$. The corresponding Lagrangian in phase space representation is given by
\begin{equation*}
(x(\tau), p(\tau)) \mapsto L((x(\tau), p(\tau)))=p(\tau) \dot{x}(\tau)-\frac{p(\tau)}{2} - \frac{1}{2}k x(\tau)^2.
\end{equation*}
In addition to the matrix $K$ from the free case, see \eqref{kinmat}, we have a matrix $L$ which includes the information about the potential, see also \cite{GS98a}. In order to realize \eqref{anpsfey} for the harmonic oscillator we consider
\begin{multline*}
I_{HO}={\rm Nexp}\big(-\frac{1}{2} \langle (\omega_x,\omega_p), K (\omega_x,\omega_p) \rangle \big) \cdot \exp\big(-\frac{1}{2} \langle (\omega_x,\omega_p), L (\omega_x,\omega_p) \rangle\big)\\
 \cdot \delta\big(\langle (\omega_x,\omega_p), (\1_{[0,t)},0) \rangle -y\big),
\end{multline*}
with
\begin{equation*}
L=\left(
\begin{array}{l l}
i k A & 0\\
0 & 0
\end{array}
\right),\, y \in \R, \,t>0.
\end{equation*}
Here $A \,f(s)=\1_{[0,t)}(s) \int_s^t \int_0^{\tau} f(r) \, dr \, d\tau, f \in L^2(\R,\C), s\in \R$.
Hence we apply Lemma \ref{thelemma} to the case
\begin{equation*}
N=\left(
\begin{array}{l l}
\1_{[0,t)^c}+ikA &-i\1_{[0,t)} \\
-i\1_{[0,t)} & \1_{[0,t)^c}+i\1_{[0,t)}
\end{array}
\right).
\end{equation*}
For determining the inverse of $N$ we use the decomposition of $L^2_2(\R)_{\C}$ into the orthogonal subspaces $L^2_2([0,t))_{\C}$ and $L^2_2([0,t)^c)_{\C}$. The operator $N$ leaves both spaces invariant and on $L^2_2([0,t)^c)$ it is already the identity. Therefore we need just an inversion of $N$ on $L^2_2([0,t))$.
By calculation we obtain 
\begin{align*}
N^{-1}= 
 \left(
\begin{array}{l l}
\1_{[0,t)^c}& 0 \\
0 & \1_{[0,t)^c} \end{array}
\right) -
\1_{[0,t)} \left(
\begin{array}{l l}
i (kA- \1_{[0,t)})^{-1} & i (kA- \1_{[0,t)})^{-1} \\
i (kA- \1_{[0,t)})^{-1}& ik A(kA- \1_{[0,t)})^{-1} \end{array}\right),
\end{align*}
if $(kA- \1_{[0,t)})^{-1}$ exists, i.e.~$kA- \1_{[0,t)}$ is bijective on $L^2_2([0,t))$. 
The operator $kAf(s) =\1_{[0,t)}(s) k \int_s^t \int_0^{\tau} f(r) \, dr \, d\tau$, $f \in L^2_2([0,t))_{\C}, s \in [0,t)$, diagonalizes and the eigenvalues $l_n$ different from zero have the form: 
\begin{equation*}
l_n=k  \bigg(\frac{t}{(n-\frac{1}{2})\pi}\bigg)^2, \quad n \in \N.
\end{equation*}
Thus $(kA- \1_{[0,t)})^{-1}$ exists if $l_n \neq 1$ for all $n\in \N$. For $0<t<\pi/(2\sqrt{k})$ this is true. 
The corresponding normalized eigenvectors to $l_n$ are 
\begin{equation*}
[0,t)\ni s \mapsto e_n(s)=\sqrt{\frac{2}{t}}\cos\left(\frac{s}{t}\left(n-\frac{1}{2}\right)\pi\right), \quad s \in [0,t) \quad n \in \N.
\end{equation*}
Hence we obtain using \cite[p.~431, form.~1]{GR65}: 
\begin{align*}
\frac{1}{\det(Id +L(Id+K)^{-1})}
&=\det\left( Id + \left(\begin{array}{l l}
-k A & -k A \\
0&0 
\end{array}
\right)\right)^{-1}\\
&= \big(\prod_{n=1}^{\infty} (1-k \big(\frac{t}{(n-\frac{1}{2})\pi}\big)^2)\big)^{-1}=\frac{1}{\cos(\sqrt{k} t)}.
\end{align*} 
Furthermore, again with $\boldsymbol{\eta}= (\1_{[0,t)},0)$ we obtain 
\begin{multline*}
({\boldsymbol\eta}, N^{-1} {\boldsymbol\eta}) = (\1_{[0,t)},({\bf \1}_{[0,t)^c}-i({\bf \1}_{[0,t)}-kA)^{-1})\,\1_{[0,t)})
=i\sum_{n=1}^{\infty} \big(1- l_n)^{-1} (\1_{[0,t)},e_n)^2 \\
=i \sum_{n=1}^{\infty} \frac{1}{1-k  \big(\frac{t}{((n-\frac{1}{2})\pi}\big)^2} \frac{2t}{(n-\frac{1}{2})\pi)^2} 
=2i t \sum_{n=1}^{\infty} \frac{1}{((n-\frac{1}{2})\pi)^2 - k  t^2 }\\
=\frac{i}{\sqrt{k}}   8 \sqrt{k}t \sum_{n=1}^{\infty} \frac{1}{((2n-1)\pi)^2 - 4k t^2 }=\frac{i}{\sqrt{k}}  \tan(\sqrt{k}t) =i \frac{\tan(\sqrt{k} t)}{\sqrt{k}},
\end{multline*}
by using \cite[p.~421,form.~1]{GR65}. Hence we have for the $T$-transform in ${\bf f} \in S_2(\R)$ by applying Lemma \ref{thelemma}  
\begin{multline}\label{genfunho}
T I_{HO}({\bf f})= \sqrt{\left(\frac{\sqrt{k}}{2\pi i \sin(\sqrt{k} t)}\right)} \exp\!\left(- \frac{1}{2} \frac{\sqrt{k}}{i\tan(\sqrt{k} t)} \Big(y-\big({\boldsymbol{\eta}}, {\bf f} +{\bf g}\big) \Big)^2\right)\\
\times\exp\!\Bigg(-\frac{1}{2} \bigg( \big({\bf f} + {\bf g}\big) ,\! 
 \left(
\begin{array}{l l}
\1_{[0,t)^c} & 0 \\
& \1_{[0,t)^c} \end{array}\right)
\big({\bf f} + {\bf g}\big) \bigg)\!\Bigg)\\
\times\exp\!\Bigg(-\frac{1}{2} \bigg( \big({\bf f} + {\bf g}\big) ,\! 
 \left(
\begin{array}{l l}
- i\1_{[0,t)} (kA- \1_{[0,t)})^{-1} & -i \1_{[0,t)}(kA- \1_{[0,t)})^{-1} \\
-i \1_{[0,t)}(kA- \1_{[0,t)})^{-1}& -ik \1_{[0,t)}A(kA- \1_{[0,t)})^{-1} \end{array}\right)
\big({\bf f} + {\bf g}\big) \bigg)\!\Bigg)
\end{multline}
Summarizing we have the following theorem: 
\begin{theorem}\label{hothm}
Let $y\in \R$, $0<t<\frac{\pi}{2\sqrt{k}}$, then the Feynman integrand for the harmonic oscillator in phase space $I_{H0}$ exists as a Hida distribution and its generating functional is given by \eqref{genfunho}. Moreover its generalized expectation
\begin{equation*}
\mathbb{E}(I_{HO})=T(I_{HO})(0)=\sqrt{\left(\frac{\sqrt{k}}{2\pi i \sin(\sqrt{k} t)}\right)} \exp\left( i \frac{\sqrt{k}}{2\tan(\sqrt{k} t)} y^2\right)
\end{equation*}
is the Greens function to the Schrö\-dinger equation for the harmo\-nic oscil\-lator, compare e.g.~with \cite{KL85}.   
\end{theorem}

\section{Canonical commutation relations}
In this section we give a functional form of the quantum mechanical commutator relations. The definition can be found in \cite{FeHi65}, for their realization in the white noise framework, we refer to \cite[Chap.~9]{W95}. With the help of these relations we can confirm that the choice of the phase space variables, as in \eqref{varchoice}, gives the right physics. I.e.~the variables fulfill the non-commutativity of momentum and position variables at equal times. This seemed to have no direct translation in a path integral formulation of quantum mechanics. But on a heuristic level Feynman and Hibbs \cite{FeHi65} found an argument to show that $\mathbb{E}(p(t+\varepsilon) x(t)I_V) \neq \mathbb{E}(p(t-\varepsilon) x(t)I_V)$ for infinitesimal
small $\varepsilon$ and that the difference is given by the commutator. First we collect some helpful formulas. 
\begin{lemma}\label{Wickwhite}
Let $\Phi \in (S)'$, ${\bf k} \in S_d(\R)$ and $n\in \N$, then 
\begin{equation*}
(-i)^n \frac{d^n}{d \lambda^n} T \Phi (\lambda {\bf k} + {\bf f})_{|\lambda =0} = T(\langle {\bf{k}}, \cdot \rangle^n \cdot \Phi)({\bf f}), \quad {\bf f} \in S_d(\R).
\end{equation*}
\end{lemma}
The proof of this lemma is an easy application of Corollary \ref{seqcor}. Note that for $\Phi \in (S)'$, ${\bf k} \in S_d(\R)$, $n \in\N$ the product $\langle {\bf{k}}, \cdot \rangle^n \cdot \Phi$ in Lemma \ref{Wickwhite} is defined by using that $(S)$ is a continuous algebra w.r.t.~the pointwise product.
In the following for ${\boldsymbol\eta_i},{\bf k} \in L^2_d(\R)$ and $y_i \in \R$, $i \in (1,\dots,J)$, we use the abbreviations: $\langle {\boldsymbol\eta}, N^{-1}{\bf k} \rangle=\big(
({\boldsymbol\eta_1},N^{-1}{\bf k}),
\dots,
({\boldsymbol\eta}_J,N^{-1}{\bf k}) 
\big)\in \R^J$ and $y=(y_1,\dots, y_J)\in \R^J$.

\begin{proposition}\label{commlemm}
Let $\Phi_{K,L}$ be as in Lemma \ref{thelemma}. Then for ${\bf k}, {\bf h} \in L^2_2(\R)$ $\langle{\bf k}, \cdot \rangle\cdot \Phi_{K,L}$ and $\langle {\bf h}, \cdot \rangle \cdot \langle{\bf k}, \cdot \rangle \cdot \Phi_{K,L}$ exist as Hida distributions. Furthermore for ${\bf f} \in S_d(\R)$
\vspace*{-0.5 cm}
\begin{multline*}
T(\langle{\bf k}, \cdot \rangle \cdot\Phi_{K,L})({\bf f}) \\
= i T\Phi_{K,L}({\bf f})
\Bigg( \Big( {\bf f} , N^{-1} {\bf k}\Big)
+\Big( \langle {\boldsymbol\eta}, N^{-1}{\bf k} \rangle, 
M_{N^{-1}}^{-1} 
\Big( i y+ \langle {\boldsymbol\eta}, N^{-1}{\bf f+g} \rangle
\Big)\! \bigg)
\end{multline*}
and
\begin{multline*}
T(\langle {\bf k}, \cdot \rangle \cdot \langle {\bf h}, \cdot \rangle \cdot \Phi_{K,L})({\bf f})
= T(\Phi_{K,L})({\bf f})
 \Bigg(\bigg(({\bf k},N^{-1}{\bf h}) +\bigg( \langle {\boldsymbol\eta}, N^{-1}{\bf h} \rangle, M_{N^{-1}}^{-1}  \langle {\boldsymbol\eta}, N^{-1}{\bf k} \rangle 
\bigg) \bigg) \\
-\bigg(\big(({\bf f} + {\bf g}),N^{-1}{\bf h}\big) +\bigg( \Big( i y+ \langle {\boldsymbol\eta}, N^{-1}{\bf f+g} \rangle\Big), M_{N^{-1}}^{-1}  \langle {\boldsymbol\eta}, N^{-1}{\bf h} \rangle  \bigg) \bigg)\\
\times \bigg(\big(({\bf f} + {\bf g}),N^{-1} {\bf k}\big) +\bigg( \Big( i y+ \langle {\boldsymbol\eta}, N^{-1}{\bf f+g} \rangle\Big), M_{N^{-1}}^{-1}  \langle {\boldsymbol\eta}, N^{-1}{\bf k} \rangle  \bigg) \bigg)\Bigg).
\end{multline*}
\end{proposition}
\noindent {\bf Proof:}
We have from Lemma \ref{Wickwhite} that
$T(\langle {\bf k}, \cdot \rangle \cdot \Phi_{K,L})({\bf f})= \frac{1}{i}\frac{d}{d \lambda} T(\Phi_{K,L})({\bf f}+\lambda {\bf k})_{|_{\lambda =0}}$, ${\bf k} \in S_d(\R)$.
Then by Lemma \ref{thelemma},
\begin{multline*}
T(\Phi_{K,L})({\bf f}+\lambda {\bf k}) = T(\Phi_{K,L})({\bf f}) \exp\Big(- \frac{1}{2} \lambda^2 \big( {\bf k}, N^{-1} {\bf k} \big) - \lambda \big( {\bf f}, N^{-1} {\bf k} \big) \Big) \\\exp\bigg( -\frac{1}{2} \lambda^2 \Big( 
  \langle {\boldsymbol\eta}, N^{-1}{\bf k} \rangle, M_{N^{-1}}^{-1} \langle {\boldsymbol\eta}, N^{-1}{\bf k} \rangle \Big) \\
 - \lambda \Big( \langle {\boldsymbol\eta}, N^{-1}{\bf k} \rangle,M_{N^{-1}}^{-1} 
\Big( i y+ \langle {\boldsymbol\eta}, N^{-1}{\bf f+g} \rangle
\Big)\Big)\bigg). 
\end{multline*}
Thus, by the above formula we get
\begin{multline*}
\frac{1}{i}\frac{d}{d \lambda} T(\Phi_{K,L})({\bf f}+\lambda {\bf k}) =-i T \Phi_{K,L}({\bf f}) \\
\times\bigg( -\Big( \Big( {\bf f} , N^{-1} {\bf k}\Big)
+\Big( \langle {\boldsymbol\eta}, N^{-1}{\bf k} \rangle, 
M_{N^{-1}}^{-1} 
\Big( i y+ \langle {\boldsymbol\eta}, N^{-1}{\bf f+g} \rangle
\Big) \Big) \Big) \\
- \lambda \Big( \big( {\bf k} , N^{-1} {\bf k}\big)
+\big( \langle {\boldsymbol\eta}, N^{-1}{\bf k}\rangle, 
M_{N^{-1}}^{-1} 
\langle {\boldsymbol\eta}, N^{-1}{\bf k}\rangle\big)\Big)\bigg).
\end{multline*}
Then by an approximation in the sense of Corollary \ref{seqcor} we get $\langle {\bf k}, \cdot \rangle \cdot \Phi_{K,L} \in (S)'$ for ${\bf k} \in L^2_d(\R)$. Setting $\lambda = 0$ we obtain the desired expression. In an analogue way one can show the second formula by using the second derivative, see Lemma \ref{Wickwhite} and polarization identity. $\hfill \blacksquare$\\

Next we extend this to the case, where just one of the functions is in $L^2_2(\R)$, but the other one is a tempered distribution. 
\begin{definition}
Let ${\bf h} \in L^2_d(\R)$ and ${\bf k} \in S_d'(\R)$ with compact support and let $ (\psi_{n})_{n \in \N}$ be a standard approximate identity. Since the convolution of a compactly supported smooth function with a compactly supported tempered distribution gives a Schwartz test function, i.e.~ ${\psi}_n \ast {\bf k}\in S_d(\R)$, $n\in \N$, see e.g.~\cite[Chap.9]{RS75b} we may define
\begin{equation*}
\langle {\bf k}, \cdot \rangle \cdot \langle {\bf h}, \cdot \rangle\cdot \Phi_{K,L} := \lim_{n \to \infty} \langle {\psi}_n \ast {\bf k} , \cdot \rangle\cdot \langle {\bf h} , \cdot \rangle \cdot \Phi_{K,L},
\end{equation*}
in the case the limit exists in the sense of Corollary \ref{seqcor}.
\end{definition}
In the following for convenience we restrict ourselves to the case $d=2$.  For the free Feynman Integrand we have then as an analogue to \cite{W95}:
\begin{theorem}\label{commthm}
Let $0<s-\varepsilon<s<s+\varepsilon<t<\infty$, then 
\begin{eqnarray*}
\langle \big(0,\delta_{s \pm \varepsilon}\big) , \cdot \rangle \cdot \langle \big(\1_{0,s)},0) , \cdot \rangle \cdot I_0 \in (S)'
\end{eqnarray*} and
\begin{eqnarray*}
\lim_{\varepsilon \to 0} \Big(T(\langle \delta_{s + \varepsilon} , \cdot \rangle \cdot \langle \1_{[0,s)} , \cdot \rangle \cdot I_0 ) )({\bf 0}) -T(\langle \delta_{s -\varepsilon} , \cdot \rangle \cdot \langle \1_{[0,s)} , \cdot \rangle \cdot I_0 ) )({\bf 0})\Big)
=-i T(I_0)({\bf 0}).
\end{eqnarray*}

\end{theorem}

\noindent {\bf Proof:} Set $\boldsymbol{\psi}^{\pm}_n:= \psi_n\ast \Big(0,\delta_{s\pm \varepsilon}\Big)$, $n\in \N$, where 
$(\psi_n)_{n\in \N}$ is a standard approximate identity.
Note that $\lim\limits_{n \to \infty} \langle \boldsymbol{\psi}^{\pm}_n, (0,\1_{[0,s)}) \rangle = \frac{1}{2} \pm \frac{1}{2}$. Using  \eqref{genfun} in the case $\boldsymbol{\eta}= \big(\1_{[0,t)},0\big)$ with $N^{-1}=
\bigg(
\begin{array}{l l}
 \1_{[0,t)^c}+i \1_{[0,t)}&i \1_{[0,t)}\\
    i \1_{[0,t)}& \1_{[0,t)^c}
\end{array} 
\bigg)$ 
as in \eqref{InvN},
we have $(M_{N^-1})^{-1}=\frac{1}{it}$. Thus, together with Proposition \ref{commlemm} we obtain 
\begin{multline}\label{comT}
T(\langle \boldsymbol{\psi}^{\pm}_n, \cdot \rangle\cdot \langle (\1_{[0,s)},0) , \cdot \rangle \cdot I_0)({\bf f})=T(I_0)({\bf f})\\
\times\Bigg(\!
\bigg(\!\Big(\Psi^{\pm}_n,(i\1_{[0,s)}, i\1_{[0,s)})\Big) 
+\bigg(\!\Big( (i\1_{[0,t)}, i\1_{[0,t)}),(\1_{[0,s)},0) \Big), \frac{1}{it} \Big( (i\1_{[0,t)}, i\1_{[0,t)}),\Psi^{\pm}_n) \Big) \!\bigg) \!\bigg) \\
-\bigg(
\bigg(
\Big(
{\bf f},(i\1_{[0,s)}, i\1_{[0,s)})
\Big) 
+\Big(
\big(i y + (i\1_{[0,t)}, i\1_{[0,t)}),{\bf f}\big), \frac{1}{it} \Big( (i\1_{[0,t)}, i\1_{[0,t)}),(\1_{[0,s)},0) \Big)   
\Big)
\bigg) 
\\
\times 
\bigg(
\Big(N^{-1}{\bf f}, \Psi^{\pm}_n\Big) 
+\Big(\big(
i y + ((i\1_{[0,t)}, i\1_{[0,t)}),{\bf f} )\big), \frac{1}{it} \Big((i\1_{[0,t)}, i\1_{[0,t)}),\Psi^{\pm}_n\Big) \Big)  \bigg) 
\Bigg)
\end{multline}
Now let us take a look at the terms which include the sequence $\Psi^{\pm}_n$.\\
Since $N^{-1}$ consists of projections on $[0,t)$ or $[0,t)^c$ respectively and $\int_{\R} \Psi^{\pm}_n(s) \, ds =1$, we have
$|(N^{-1}{\bf f}, \Psi^{\pm}_n)|\leq \|{\bf f}\|_{\sup} $.
Furthermore $|((i\1_{[0,u)}, i\1_{[0,u)}),\Psi^{\pm}_n)|\leq 1$, for all $n\in \N$ and $0< u\leq t$.
Therefore the expression can be bounded uniformly in $n \in \N$ in the sense of Corollary \ref{seqcor} (note that $\|\cdot\|_{\sup} \leq \|\cdot\|_p$ for some $p\in \N$). Obviously the $T$-transform in \eqref{comT} is convergent as $n\to\infty$, thus the limit exists as a Hida distribution by Corollary \ref{seqcor}. Taking the limit leads us to 
\begin{multline*}
T(I_0)({\bf f})\bigg(\frac{s}{i t}+ i\1_{[0,s)}(s\pm \varepsilon) -\bigg( \frac{i}{t} \big(s\cdot(y+ \int_{[0,t)} f_x+f_p \, ds) \big) \\
- \left(\!i\int_{0}^s f_x + f_p \,dt \right) \bigg) \bigg(\frac{i}{t} \big(y+ \int_{[0,t)} f_x+f_p \, ds \big) - \left(\!f_x (s\pm\varepsilon)\right) \bigg) \bigg),\quad {\bf f}=(f_x,f_p) \in S_2(\R).
\end{multline*}
For the difference $\mathbb{E}(\langle \delta_{s + \varepsilon} , \cdot \rangle \langle \1_{[0,s)} , \cdot \rangle I_0) - \mathbb{E}(\langle \delta_{s - \varepsilon} , \cdot \rangle \langle \1_{[0,s)} , \cdot \rangle I_0)$  we have 
\begin{multline*}
\lim_{\varepsilon \to 0}T(\langle \delta_{s + \varepsilon} , \cdot \rangle \langle \1_{[0,s)} , \cdot \rangle I_0 ) )({\bf 0}) -T(\langle \delta_{s -\varepsilon} , \cdot \rangle \langle \1_{[0,s)} , \cdot \rangle I_0 ) )({\bf 0})\\
=\lim_{\varepsilon \to 0} T(I_0)({\bf 0}) \big( i \1_{[0,s)}(s+ \varepsilon) -  i\1_{[0,s)}(s- \varepsilon)\big)
= T(I_0)({\bf 0}) \cdot ( 0 - i )
=-i T(I_0)({\bf 0}),
\end{multline*}
which completes the proof.
$\hfill \blacksquare$\\
Thus, the commutation law for the free Feynman integrand in phase space is fulfilled in the sense of Feynman and Hibbs \cite{FeHi65}. 

\noindent \underline{Acknowledgements}\\[0.5 cm]
We dedicate this article to Anatolij Skorohod, Volodymyr Korolyuk and Igor Kovalenko. The authors would like to thank the organizing and programme committee of the MSTAII conference for an interesting an stimulating meeting. Wolfgang Bock wants especially thank to Yuri Kondratiev for the opportunity to give a talk on this topic
at the conference. Furthermore the authors would like to thank Florian Conrad, Anna Hoffmann, Tobias Kuna and Ludwig Streit for helpful discussions. The financial support from the DFG project GR 1809/9-1, which enabled the authors to join the conference, is thankfully acknowledged.

\end{document}